\documentclass{article}

\usepackage{xcolor}
\usepackage{arxiv}

\usepackage[utf8]{inputenc} 
\usepackage[T1]{fontenc}    
\usepackage{hyperref}       
\usepackage{url}            
\usepackage{booktabs}       
\usepackage{amsfonts}       
\usepackage{nicefrac}       
\usepackage{microtype}      
\usepackage{lipsum}
\usepackage{graphicx}
\usepackage{CJKutf8}
\graphicspath{ {./images/} }
\usepackage{amsmath,amsthm,amssymb,amsfonts,amscd,keyval}
\usepackage{bm}
\usepackage{mathtools}
\usepackage{multirow}
\usepackage{dsfont}
\usepackage{float}
\usepackage{xspace}
\usepackage{subfig}
\usepackage{bbding}
\usepackage{makecell}
 \usepackage{threeparttable}
 \usepackage[normalem]{ulem}
 \usepackage{multirow}
\useunder{\uline}{\ul}{}
\title{BrainNPT: Pre-training Transformer Networks for Brain Functional Network Classification}

\author{%
Jinlong Hu$^1$\thanks{Corresponding author. Email: jlhu@scut.edu.cn} \quad Yangmin Huang$^1$ \quad Nan Wang$^2$ \quad Shoubin Dong$^1$ \quad \\
$^1$ Guangdong Key Lab of Communication and Computer Network\\
School of Computer Science and Engineering\\ South China University of Technology, Guangzhou, China\\
 $^2$ School of Computer Science and Technology\\
 East China Normal University, Shanghai, China\\
}

\begin{document}
\begin{CJK}{UTF8}{gbsn}
\maketitle
\begin{abstract}
Deep learning methods have advanced quickly in brain imaging analysis over the past few years, but they are usually restricted by the limited labeled data. Pre-trained model on unlabeled data has presented promising improvement in feature learning in many domains, including natural language processing and computer vision. However, this technique is under-explored in brain network analysis. In this paper, we focused on pre-training methods with Transformer networks to leverage existing unlabeled data for brain functional network classification. First, we proposed a Transformer-based neural network, named as BrainNPT, for brain functional network classification. The proposed method leveraged <cls> token as a classification embedding vector for the Transformer model to effectively capture the representation of brain network. Second, we proposed a pre-training framework for BrainNPT model to leverage unlabeled brain network data to learn the structure information of brain networks. The results of classification experiments demonstrated the BrainNPT model without pre-training achieved the best performance with the state-of-the-art models, and the BrainNPT model with pre-training strongly outperformed the state-of-the-art models. The pre-training BrainNPT model improved 8.75\% of accuracy compared with the model without pre-training. We further compared the pre-training strategies, analyzed the influence of the parameters of the model, and interpreted the trained model. 
\end{abstract}


\section{Introduction}
Deep learning methods have advanced quickly in brain imaging analysis over the past few years~\cite{ml2019,yin2022deep}. To characterize brain functional connectivity and their variability from various brain disorders, a growing number of graph deep learning models have been developed for brain functional network analysis, such as graph neural networks (GNNs) with specially designed convolutional layers~\cite{li2021braingnn}, graph attention networks~\cite{hu2021gat}, Graph isomorphism networks (GIN) combined with infomax regularizations~\cite{kim2020understanding}, GroupINN~\cite{yan2019groupinn},  IBGNN~\cite{cui2022interpretable}, and FBNetGEN~\cite{kan2022fbnetgen}. 

In recent years, Transformer~\cite{vaswani2017attention} has shown extraordinary performance in natural language processing (NLP), such as BERT~\cite{devlin2018bert} and GPT~\cite{radford2018improving}, with their multi-headed attention mechanism and the capacity to capture long-range dependency. Some studies have also tried to apply Transformer structures to the graph data domain. They usually exploited a designed position embedding strategy for specific graphs to integrate position information into node embedding for self-attention calculation. For example, Graph-Bert~\cite{zhang2020graph} proposed three position encoding strategies based on sub-graph structure, including Weisfeiler-Lehman absolute role embedding, intimacy-based relative positional embedding, and hop-based relative distance embedding. GROVER~\cite{rong2020self} utilized GNNs instead of multi-layer perception (MLP) to attain representation vectors of query, key, and value for self-attention. GraphiT~\cite{mialon2021graphit} employed GNNs as the kernel function for relative position embedding to calculate self-attention scores further. Graphormer~\cite{ying2021transformers} proposed three position encoding strategies specially designed for molecular graphs structure. Structure-aware Transformer (SAT)~\cite{chen2022structure} applied subgraph structure as position embedding. GraphTrans~\cite{wu2021representing} exploited GNNs to introduce position and local structure information. GraphGPS~\cite{rampavsek2022recipe} adapted alternately combined structure of message-passing graph neural networks (MPNNs) and Transformer to enhance position information extraction. Recently, BrainNetTF~\cite{kan2022brain} combined Transformer structure with the properties of the adjacency matrix of brain network, and achieved state-of-the-art performance for brain network classification tasks.

However, deep learning methods are usually restricted by the limited labeled data. Since collecting resting-state functional magnetic resonance imaging (rs-fMRI) data is typically time-consuming and requires laborious experiments with specialized equipment, and some brain disorders are rare and the related rs-fMRI data is even more minor scales. The data of brain functional networks constructed from rs-fMRI data is generally scarce. Besides, Transformer with insufficient data could lead to poor performance compared to traditional convolutional methods~\cite{hassani2021escaping,chen2021empirical}. Pre-training is a promising technique to address the problem of limited labeled data, for example, a GNNs-based pre-training approach has been applied to classify brain networks using brain imaging data without relevant task labels~\cite{yang2022pre}. Moreover, Transformer-based pre-training methods have achieved great success in NLP~\cite{devlin2018bert,radford2018improving} and computer vision (CV)~\cite{dosovitskiy2020image,liu2021swin}, but this pre-training technique has not been explored for brain functional network analysis. 

In this paper, we focused on pre-training methods with Transformer networks to leverage existing unlabeled data for brain functional network classification. First, we proposed a new Transformer-based neural network, named as BrainNPT, for brain functional network classification, where a learnable <cls> token was designed as a classification embedding vector to capture the representation of brain network. Second, we proposed a framework to pre-train the BrainNPT model with large-sale unlabeled brain functional network data, where replaced brain region of interest (ROI) prediction was proposed as the pre-training strategy. The BrainNPT model was evaluated with two publicly available datasets from Autism Brain Imaging Data Exchange (ABIDE) II~\cite{di2017enhancing} and REST-meta-MDD~\cite{yan2019reduced}. The results of experiments demonstrated the basic BrainNPT model without pre-training achieved the best performance with the state-of-the-art models, and the BrainNPT model with pre-training outperformed the state-of-the-art models. Moreover, We further compared the pre-training strategies, analyzed the influence of the parameters of the model, and interpreted the fine-tuned model. 

The main contributions are summarized as follows:

(i) We proposed a novel Transformer-based neural network with <cls> token for brain functional network classification, which was a conceptually simple yet effective model to classify brain networks, and could be easily pre-trained and deeply stacked in term of Transformer style. In addition, the proposed model could be interpreted by layer-wise relevance propagation (LRP), which is one of the most prominent explanation techniques for deep neural networks.

(ii) We proposed a pre-training framework with an effective pre-training strategy, which detected whether the brain ROIs had been replaced. These pre-training strategy enabled the model to capture the intrinsic brain functional network structures from unlabeled brain network data, and the pre-training model was improved 8.75\% of accuracy compared with the basic model without pre-training.

(iii) We augmented brain functional network data by sliding windows over time on rs-fMRI data, and used a large amount of these brain snapshot network data for pre-training. Compared with the powerful capability of Transformer-based model, even unlabeled brain functional network data is scarce, and the experimental results indicated that using sliding windows for data augmentation was very helpful for pre-training the model.
\section{Methods}
\label{sec:method}
\subsection{Definition}
\label{sec:notation}
In this paper, the brain functional networks are considered as graphs, where each ROI is considered as a node in the graph, and the functional connectivity values between ROIs are considered as the edge weights between nodes. 

Based on preprocessed rs-fMRI data, $N$ ROIs are obtained according to a brain atlas for each subject, the time series of ROI is calculated by averaging the values of all voxels in the ROI, functional connectivity is obtained by Pearson correlation operation between the time series of ROIs, and a functional connectivity matrix with the shape of $N \times N$ is constructed for each subject. Regarding the functional connectivity matrix as a brain functional network $X \in {{\mathbb R}^{N \times N}}$, a Fisher transformation is also performed with the brain network for normality. In addition, we consider a vector containing functional connectivity values between a certain ROI with all other ROIs as the feature vector of that ROI.

\subsection{Architecture of BrainNPT model}
We proposed a Transformer based neural network, namely BrainNPT, which leveraged the <cls> classification embedding vector as a virtual ROI, inspired by the <cls> token in NLP, to collect rich structure patterns of brain networks for classification.

\begin{figure*}[h]
    \centering
    \includegraphics[width=0.9\linewidth]{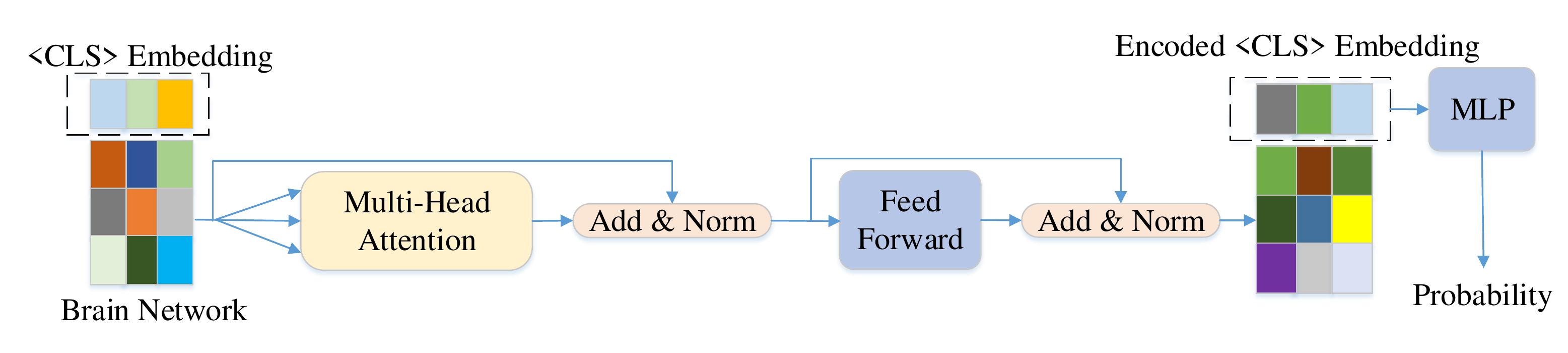}
    \caption{The architecture of BrainNPT. The architecture includes classification embedding vector <cls>, Transformer block, and MLP block. The Transformer block could be stacked into multiple layers.}
    \label{fig:model}
\end{figure*}

The architecture of our proposed BrainNPT is shown in Fig.~\ref{fig:model}. BrainNPT is constructed based on the Transformer structure~\cite{vaswani2017attention}, which includes three parts, classification embedding vector <cls>, Transformer block, and multi-layer perceptron (MLP) module. For the brain network with the input shape of $N \times N$, we firstly concatenate it with a learnable classification embedding vector <cls>, with the length of $N$, and $\left( {N{\rm{ + }}1} \right) \times N$ as the actual input shape. Afterwards, we take the embedding vector <cls> as the representation vector of the brain functional network after forwarding it into Transformer blocks. Finally, the representation vector of the brain functional network is forwarded into the MLP module for generating classification probability.

In order to exploit the positional information of the input sequence, the Transformer methods usually adopt a position embedding strategy for Transformer-based methods from NLP or CV domain, such as sin-cos encoding strategy~\cite{devlin2018bert}. For the brain functional networks, since the functional connectivity matrix is symmetry and density, connectivity values represent the connection information between ROIs, and using eigenvectors or other position encoding strategies for brain networks with density is costly and ineffective~\cite{kan2022brain}. In addition, combining edge weights into self-attention could impair the effectiveness and performance of self attention\cite{cui2022braingb}. Therefore, the functional connectivity matrix was directly regarded as the input with position embedding information for further self-attention operation in our proposed method.
\subsubsection{Classification embedding vector <cls>} 
Inspired by GraphTrans~\cite{wu2021representing} and the <cls> token in NLP, we utilized a learnable classification embedding vector <cls> for the Transformer in BrainNPT, which was distinguished from the traditional read-out layer of GNN methods such as max-pooling, average pooling, and attention-based read-out layers~\cite{lee2019self,li2015gated}. Specially, we concatenate the <cls> classification embedding vector with the brain functional network before forwarding it as the input for self-attention module, and set it as learnable parameter. Since we do not have a position embedding module, the <cls> classification embedding vector is randomly initialized with the model parameters in the initial state of training, and it has no actual meaning. And in further training, the model would implicitly learn to collect global information of the brain functional network into the <cls> embedding vector. 

In BrainNPT, the <cls> classification embedding vector could be regarded as a virtual ROI of the brain functional network. When performing self-attention operation, this virtual ROI interacts with other ROIs and acquires information from other ROIs. Meanwhile, it helps self-attention modules learn more about long-range interaction dependencies between ROIs.
\subsubsection{Transformer block} 
The Transformer block module includes two core modules of Transformer, self-attention and feed-forward neural networks (FFN). Considering the fixed ROI positions in the brain network, similar to the normalization operation performed on each position of a sentence in NLP~\cite{narang2021Transformer}, we performed layer normalization operations in each of the two sub-modules to normalize the connectivity vectors of each ROI from the input brain network, and used residual connections to ensure training stability. Moreover, we deploy GELU~\cite{hendrycks2016gaussian} as the activation function in the Transformer block to increase the model’s expressive power.

To learn the interactions between ROIs from brain networks in the Transformer block, we leveraged the bi-directional encoding strategy without additional masks, same as in BERT~\cite{devlin2018bert}, for BrainNPT.

The self-attention and FFN sub-modules of the Transformer blocks were defined as follows. For self-attention, we employ learnable weighted matrices to project the brain functional network into three representation matrices (Query, Key, and Value), and we obtain the weights of ROIs for each ROI with Query and Key. Finally, the updated embedding with Value and weights of ROIS are obtained, as shown in (1):
\begin{equation}
Attention\left( x \right) = softmax\left( {\frac{{Qx{K^T}x}}{{\sqrt {{d_k}} }}} \right)Vx
\label{eq:attention}
\end{equation}
Where $Q$, $K$, $V$ denote Query, Key, and Value, ${d_k}$ represents the dimension of connectivity vectors, and $softmax\left(  \cdot  \right)$ denotes the softmax function.

And then, a multi-head self-attention mechanism is adopted, and a concatenating strategy is adopted for the representation matrix generated by each head to collect the final representation as the output of self-attention sub-module, as shown in (2):
\begin{equation}
mh\_Attention\left( x \right) = Concatenate\left( {Attention\left( x \right)} \right)
\label{eq:mhatten}
\end{equation}
Where $Concatenate\left(  \cdot  \right)$ represents the concatenating operation.

Finally, the output of self-attention sub-module is fed into FFN for further enhancing un-linear expressivity, as shown in (3):
\begin{equation}
FFN\left( x \right) = max\left( {0,x{W_1} + {b_1}} \right){W_2} + {b_2}
\label{eq:ffn}
\end{equation}
Where ${W_1}$ and ${W_2}$ are weighted matrices for linear projection, and ${b_1}$ and $b_2$ denote biases. With the classification embedding vector <cls> acquiring global representation information, it is exploited as the representation vector of the brain network for classification.
\subsubsection{MLP}
The obtained representation vector of brain functional network is forwarded into MLP for classification, with softmax as the final layer for generating prediction probabilities. The dropout layers are used in MLP to help reduce overfitting.
\subsection{Pre-training framework for BrainNPT}
\label{sec:pre-train}
\begin{figure*}[h]
    \centering
    \includegraphics[width=0.6\linewidth]{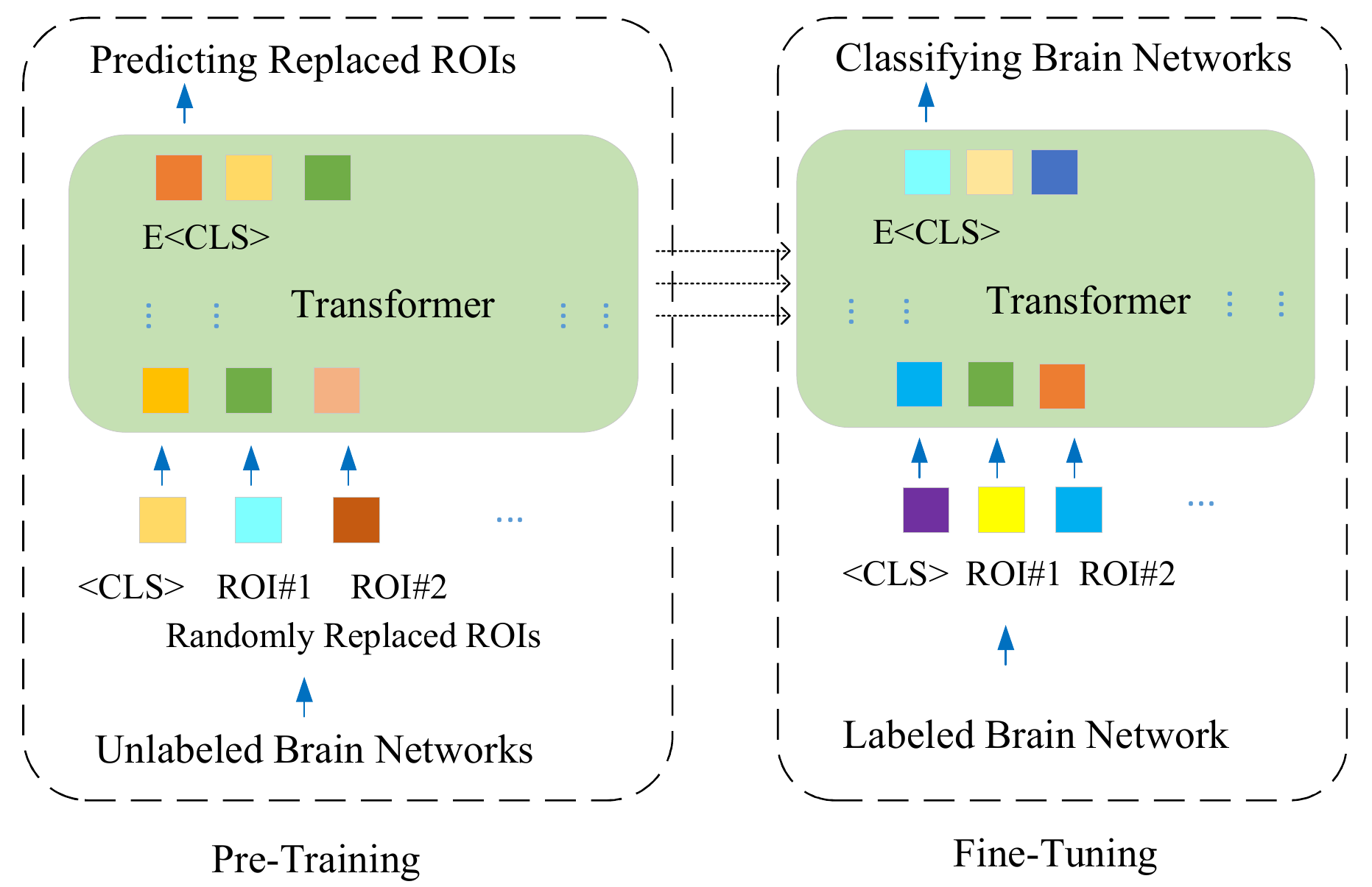}
    \caption{The framework of pre-training and fine-tuning of BrainNPT. The pre-training part used randomly replaced ROI strategy in the left block, and the fine-tuning part used the parameters from pre-training for downstream tasks in the right block.}
    \label{fig:pretrain}
\end{figure*}

We proposed a pre-training framework for BrainNPT, including the construction of pre-training datasets and the designed pre-training strategy. Before fine-tuning for downstream classification tasks, the BrainNPT is first trained with pre-training tasks to initialize parameters and learn the structural information of brain networks. The framework of BrainNPT’s pre-training and fine-tuning is shown in Fig.~\ref{fig:pretrain}. The settings of the Transformer module and <cls> classification embedding vectors are the same among pre-training models and fine-tuning models, and the difference between the pre-training models and fine-tuning models is mainly the MLP part.

In order to leverage unlabeled data for pre-training learning, we first proposed a method for constructing a brain functional network pre-training dataset. Then, we proposed the pre-training strategy for BrainNPT to acquire the general structure information of brain functional networks before fine-tuning downstream classification tasks.
\subsubsection{Construction of pre-training dataset}
In this study, we constructed pre-training datasets with sliding-window processing method to enlarge the scale of pre-training dataset. Based on unlabeled and preprocessed rs-fMRI data, we attained multiple functional connectivity matrices from each subject, with the length of windows $W$, the steps $S$, and the extracted time-series for brain networks according to the same brain atlas as the downstream labeled data. Since the connectivity matrix of each time window (the snapshot network) still reflects some characteristic of brain functional networks from the entire time period, it can be regarded as brain network data for pre-training tasks. With the sliding-windows operation, the scale of pre-training data is effectively extended. Specifically, for a subject with the length $T$ of time-series, $\left( {\left( {T - W} \right)/{\rm{S}} + 1} \right)$ brain networks can be constructed by sliding-windows for further pre-training.
\subsubsection{Replaced ROI Prediction Strategy}
\begin{figure*}[h]
    \centering
    \includegraphics[width=0.4\linewidth]{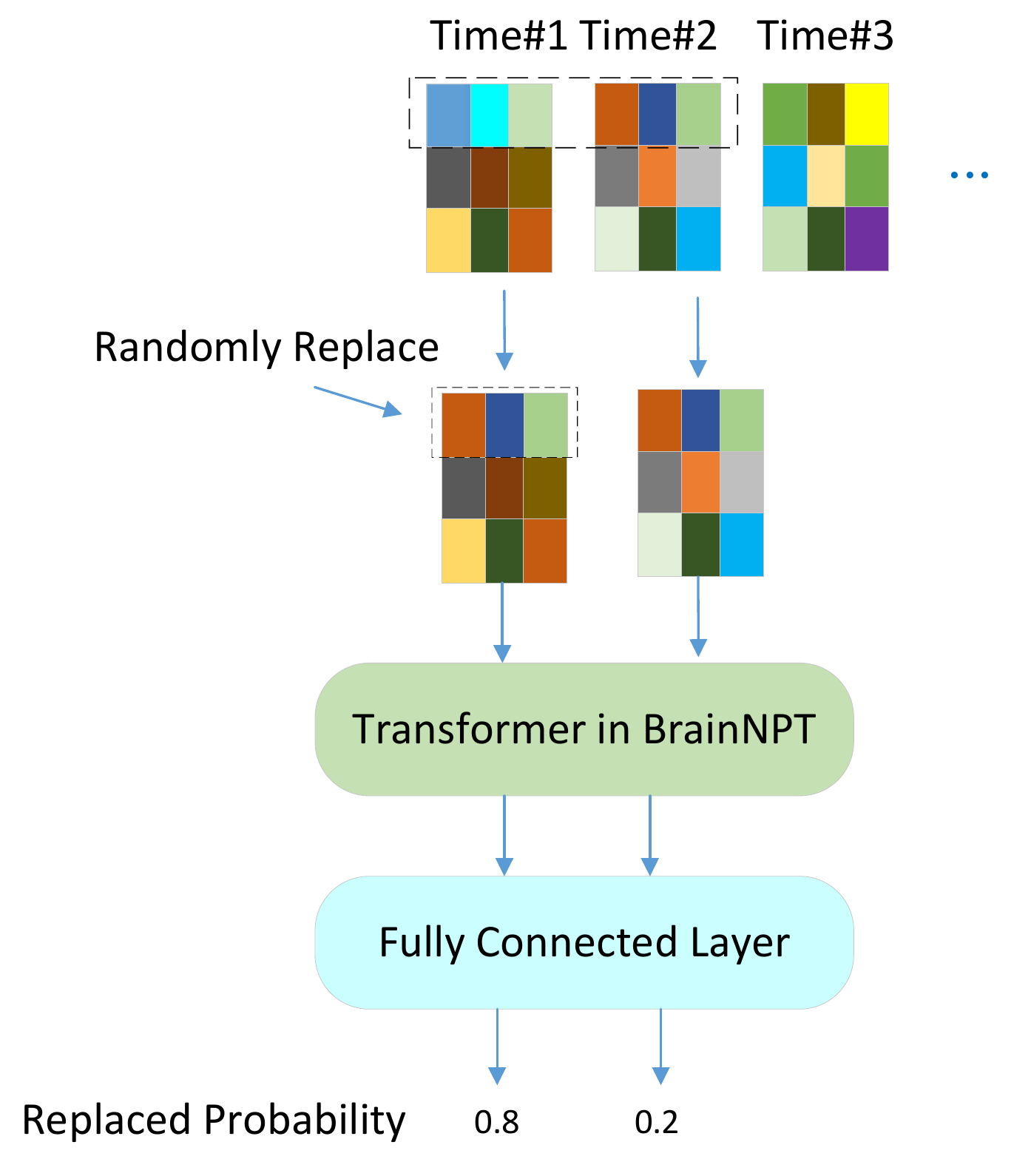}
    \caption{The framework of pre-training with RRP.}
    \label{fig:rrp}
\end{figure*}

Downstream classification tasks require models to have a comprehension of the interactions between ROIs and capture the structure information of brain networks, such as the subgraph patterns from default mode network (DMN) of the brain~\cite{raichle2015brain}. That would require models to learn the characteristic of functional connectivity matrix, such as symmetry. To enable model to acquire interactions of ROIs in brain networks, we proposed a Replaced ROI Prediction (RRP) strategy, as shown in Fig.~\ref{fig:rrp}. This strategy took a pair of brain networks as the input for the RRP model with predicting the probability of being replaced. Specifically, we took brain networks $A$ and $B$ as the input. With 50\% probability, we directly considered $A$ as the input of the Transformer part of the RRP model. With the remaining 50\% probability, we replaced 50\% of ROIs of $A$ with the functional connectivity vectors of the corresponding ROIs in $B$. In order to increase the learning difficulties, the most similar pairs of functional connectivity matrices were selected for $A$ and $B$ in experiments. We selected functional connectivity matrices of adjacent time windows from the same subject as the input $A$ and $B$. 

In the pre-training phase of the RRP model, for the binary classification tasks, we used the same settings with downstream fine-tuning tasks. The <cls> classification embedding vector was concatenated to the inputs and forwarded into the Transformer part of the RRP model. The settings help the model implicitly learn to collect global information in <cls> classification embedding vectors. Finally, we took the <cls> classification embedding vector of output of the Transformer part as the representation vector of brain networks. Using a fully connected layer with the output shape of 2, and the softmax function to predict whether the brain networks have been replaced, supervised learning was performed with binary cross-entropy loss function for the RRP model.

\subsection{Model explanation}
The Transformer models with the <cls> embedding vectors, such as BERT-base~\cite{devlin2018bert} models from NLP domain and ViT-base~\cite{dosovitskiy2020image} models from CV domain, have been demonstrated to be interpretable using self-attribution scores~\cite{hao2021self} and layer-wise relevance propagation (LRP)~\cite{chefer2021Transformer}. In this study, we adapted an LRP based interpretation method for the BrainNPT model to explore which ROIs in the brain networks have the key impact on classification.

The BrainNPT contains <cls> classification embedding vector, the Transformer block, and fully connected layers, and it is able to use LRP to calculate the relevance scores of each ROI for the classification results. Based on LRP for interpretation of BrainNPT, we could obtain the local relevance for an input sample using deep Taylor decomposition method~\cite{montavon2017explaining}. For a trained BrainNPT model, its relevance scores of ROIs can be calculated based on LRP for Transformer’s self-attention, FFN, GELU activation function and MLP. Finally, by the chain rule, the relevance score of each ROI can be estimated in the input brain network. Specifically, for interpretation output $C$ for the Transformer blocks, considering the BrainNPT only leverages <cls> classification embedding vectors as the representation vectors, we regarded the corresponding interpretation outputs ${C_{\left\langle {cls} \right\rangle }} \in {{\mathbb R}^s}$ as the interpretation results for classification, with the relevance score of each ROI to the classification results.

\section{Results and discussion}
\subsection{Classification performance of BrainNPT model}
\label{sec:basic_experiments}

\subsubsection{Experimental datasets and preprocessing}
We used two public rs-fMRI datasets, ABIDE II dataset~\cite{di2017enhancing} and REST-meta-MDD~\cite{yan2019reduced}, for our comparison classification experiments of BrainNPT. We acquired the preprocessed data of ABIDE II~\cite{craddock2013neuro}, and we selected 540 samples including 288 autistic patients and 252 healthy controls by removing some samples with poor quality. For REST-meta-MDD dataset, screened out some subjects with poor image quality, we retained a total of 2027 subjects, including 1041 patients with depression and 986 healthy controls. The data of the two datasets were both preprocessed with DPARSF~\cite{yan2010dparsf}. For each subject, we parcellated the whole brain into 200 brain regions/ROIs according to Craddock200 atlas ~\cite{craddock2013neuro}. Refer to the construction process defined in Section 3, we constructed a functional connectivity matrix of the size of $200 \times 200$ for each subject from the two datasets.

\subsubsection{Comparison to state-of-the-art models}
We compared the basic BrainNPT model without pre-training with three types of models, including typical
GNN models, GNN models designed for brain functional networks, and Transformer-based models designed for graph
data. There were twelve models from three types: (i) For typical GNN methods, we selected GAT~\cite{velivckovic2017graph}, GCN~\cite{kipf2016semi}, GraphSAGE~\cite{hamilton2017inductive}, GIN~\cite{xu2018powerful} and DiffPool~\cite{ying2018hierarchical}, which were proposed for capturing graph structure information and used for graph data from various areas. Specifically, for GAT, GCN and GraphSAGE, global attention layers~\cite{li2015gated} were selected as the read-out layer. Moreover, for typical GNNs methods, we constructed the brain networks with edge binarization, setting the threshold with 0.5 of connectivity values. (ii) For GNN methods designed for brain networks, we selected BrainGNN~\cite{li2021braingnn} BrainnetCNN~\cite{kawahara2017brainnetcnn} and FbNet~\cite{kan2022fbnetgen} as the compared methods. They were proposed for brain networks for leveraging the characteristic of brain networks, such as time-varying and relationship between ROIs. (iii) For Transformer-based methods designed for graph data, we compared our proposed methods with Transformer-based methods for graph data including brain networks, consisting of BrainNetTF~\cite{kan2022brain}, SAN~\cite{kreuzer2021rethinking}, GraphGPS~\cite{rampavsek2022recipe}, and GraphTrans~\cite{wu2021representing}. GraphTrans has specified designs of positional embedding strategies for graph data.

For our proposed BrainNPT, we adapted five layers of Transformer, three fully connected layers. The dropout rate of Transformer was set as 0.1, the dimension of FFN was set as 800, heads of self-attention as 5, and GELU as the activation function, dropout rate was 0.5 for dropout layers with fully connected layers.

Moreover, AdamW~\cite{loshchilov2017decoupled} was selected as the optimizer, and linear increasing warm-up and inverse square root decay strategy was selected as the scheduler for the learning rate~\cite{raffel2020exploring} during training. The learning rate was set as 1, with binary cross-entropy loss and 500 epochs for training. In experiments, we randomly split 80\% of the dataset for training, 10\% for validation and 10\% for testing. The performance are evaluated based on the average of 5 random runs on the test set with the standard deviation.

In the experiments, accuracy and AUC were selected for the metric of performance of classification.
We compared the BrainNPT model without pre-training with twelve models on ABIDE II dataset and rest-meta-MDD dataset, and the classification results are shown in Table~\ref{tab:basic}. 

\begin{table}[h]
\centering
\caption{The classification performance on ABIDE II and REST-meta-MDD ($\%$, mean±std).
}
\label{tab:basic}
\setlength{\tabcolsep}{30pt}{
\resizebox{\linewidth}{!}{
\begin{threeparttable}
\begin{tabular}{ccccccc}
\hline
\multirow{2}{*}{Type}                                      & \multirow{2}{*}{Method}   & \multicolumn{2}{c}{ABIDE II}        & \multicolumn{2}{c}{REST-meta-MDD}               \\ \cline{3-7} 
                                                           &                           & Accuracy            & AUC              & Accuracy            & \multicolumn{2}{c}{AUC}                \\ \hline
\multirow{5}{*}{Typical GNNs}          & GAT         & 57.41±6.62          & 59.21±7.68   & 64.53±2.51          & 67.95±4.03         \\
                                        & GCN         & 61.11±7.67          & 66.92±7.79  & 64.73±2.16          & \textbf{69.11±2.97}        \\
                                        & GraphSAGE   & 58.33±4.67          & 63.69±4.83   & 62.23±2.17          & 66.00±2.05       \\
                                        & GIN         & 62.00±6.89          & 60.73±7.60   & 56.09±3.40          & 55.48±4.10        \\
                                        & DiffPool    & 59.26±6.88          & 59.90±7.23   & 64.05±2.68          & 68.87±4.61       \\ \hline
\multirow{3}{*}{\makecell{GNN for \\ Brain Networks}} & BrainGNN    & 59.26±6.76          & 66.19±12.52  & 59.80±2.59          & 63.79±5.32        \\
                                        & BrainnetCNN & 60.69±3.23          & 64.75±2.88  & 60.99±1.44          & 66.16±1.29           \\
                                        & FBNet       & 58.78±4.01          & 62.90±5.00  & 63.05±2.16          & 65.52±0.90         \\ \hline
\multirow{3}{*}{Graph Transformer}            & SAN         & \textbf{62.03±10.95}         & 61.14±14.64   & 61.25±2.33         & 63.71±2.83        \\
                                        & GraphGPS    & 59.88±7.48          & 57.21±7.78   & \textbf{65.68±2.22}          & \textbf{69.73±1.97}         \\
                                        & GraphTrans  & 61.11±8.25          & 62.60±9.82     & 63.25±2.54          & 69.01±2.94       \\
                                        & BrainNetTF  & 61.21±5.12          & \textbf{67.60±5.69}   & 62.76±1.31          & 68.27±1.82        \\                                        \hline
Ours                                    & BrainNPT    & \textbf{62.50±7.14} & \textbf{67.88±6.80}  & \textbf{65.42±3.21} & 68.46±3.17\\ \hline
\end{tabular}
\begin{tablenotes} 
\item In each row, the top two highest accuracy and AUC are highlighted in bold. 
\end{tablenotes} 
\end{threeparttable} 
}}
\end{table}
From Table~\ref{tab:basic}, our proposed BrainNPT model achieved the best performance among the thirteen models, with the highest accuracy and AUC in ABIDE II dataset and the top two of accuracy in REST-meta-MDD dataset. Compared with other models, Transformer-based models achieved competitive performance, where BrainNetTF achieved the top two of AUC in ABIDE II dataset, and GraphGPS achieved the highest accuracy and AUC in REST-meta-MDD dataset. That indicated the power of Transformer for brain network classification. In summary, the BrainNPT model without pre-training achieved best performance among two datasets, and Transformer-based models have shown excellent capabilities for brain network classification.

\subsubsection{BrainNPT model analysis}
\textbf{Read-out layer analysis}. For graph classiﬁcation, the read-out layer usually aggregates node features to obtain the whole-graph representation. We compared <cls> classification embedding vectors with two types of read-out layer, orthonormal clustering read-out layer~\cite{kan2022brain} and average pooling read-out layer~\cite{kim2021learning} in BrainNPT model. 
The results of different read-out layers with BrainNPT on ABIDE II  are shown in Table~\ref{tab:readout}.
\begin{table}[h]
\centering
\caption{Different read-out layers in BrainNPT (mean±std).
}
\label{tab:readout}
\begin{tabular}{cccc}
\hline
Read-out                                   & Accuracy (\%)                & AUC (\%)                     \\ \hline
\multicolumn{1}{c}{orthonormal clustering} & 61.21±5.12          & 67.60±5.69\\
\multicolumn{1}{c}{average polling}        & 58.80±4.68          & 57.36±4.31 \\ 
Ours                                       & \textbf{62.50±7.14} & \textbf{67.88±6.80} \\
\hline
\end{tabular}
\end{table}

From the Table~\ref{tab:readout}, our proposed <cls> vector achieved the best performance in term of Accuracy and AUC. The orthonormal clustering achieved the similar performance compared to our proposed method but slightly lower. That might be caused by the complexity of orthonormal clustering and the small scale of dataset. 

\textbf{The influence of layers of Transformer blocks}. Furthermore, experiments on the influence of the number of layers of Transformer blocks in BrainNPT were conducted on ABIDE II, as shown in Table~\ref{tab:layer}.

\begin{table}[h]
\centering
\caption{Different layers of Transformer blocks in BrainNPT (mean±std).}
\label{tab:layer}
\begin{tabular}{ccc}
\hline
Layers & Accuracy (\%)                & AUC (\%)                      \\ \hline
2      & 58.33±8.50          & 57.41±6.61                 \\   
3      & 57.29±2.50          & 57.31±4.17                 \\
4      & 60.42±5.52         & 67.01±9.04             \\
5      & 62.50±7.14 & \textbf{67.88±6.80}         \\
6      & \textbf{62.50±6.24}         & 64.35±5.05         \\ \hline
\end{tabular}
\end{table}
The model with 5 layers of Transformer blocks achieved the best performance of the accuracy and AUC. Our model performance continued to increase as the number of layers increased from 2 to 5, but the performance of 6 layers did not increase significantly from 5 layers. The expressivity and performance of models might be restricted by the scale of datasets with deeper of Transformer blocks.
\subsection{Pre-training of BrainNPT model}
\subsubsection{Experimental datasets and preprocessing}
For pre-training experiments, in addition to ABIDE II and REST-meta-MDD dataset, 
we used two additional large-scale public datasets, Human Connectome Project (HCP) datasets~\cite{van2013wu} and ABIDE I dataset~\cite{di2014autism}. 

For ABIDE I dataset, we used the rs-fMRI data of 974 subjects selected with high image quality~\cite{craddock2013neuro}, including 467 autistic patients and 507 healthy controls. The rs-fMRI data from  ABIDE I have been preprocessed by DPARSF~\cite{yan2010dparsf}. For HCP dataset, we used the S1200 dataset, with removing some subjects with pseudo-noise in collected images, and retained 1088 subjects. The data of HCP dataset were preprocessed by GRETNA~\cite{wang2015gretna} with head motion correction, regularization, and smoothing. For each subject, 200 brain regions/ROIs were defined by Craddock200 atlas.

Refer to the sliding-window process defined in Section 4.2, the window size was fixed to 30, and the step size of sliding window was set based on the scanning parameters, respectively. Afterwards, we performed the Pearson correlation coefficient operation between the time series of ROIs within each time window to construct the functional connectivity matrix, and used Fisher transformation on the matrix. The settings of sliding-window of ABIDE I, REST-meta-MDD, and HCP datasets are shown in Table~\ref{tab:pretrain_dataset}.

\begin{table}[h]
\centering
\caption{Settings of sliding-windows of pre-training brain networks.}
\label{tab:pretrain_dataset}
\resizebox{\linewidth}{!}{
\begin{tabular}{cccccc}
\hline
Dataset       & Subjects & Length of Time Windows & Steps & Minimum of Time Windows & \multicolumn{1}{l}{Maximum of Time Windows} \\ \hline
ABIDE I       & 974      & 30                     & 1     & 48                      & 286                    \\
REST-meta-MDD & 2027     & 30                     & 1     & 90                      & 240                    \\
HCP           & 1088     & 30                     & 3     & 295                     & 387                    \\ \hline
\end{tabular}
}
\end{table}
We acquired 112,879 brain networks from ABIDE I dataset, 421,129 brain networks from HCP dataset, and 504,558 brain networks from REST-meta-MDD dataset for pre-training, regarding each connectivity matrix of each time window as a brain network. From these large scale of the pre-training data, we randomly selected 99\% of the data for training, and 1\% for evaluation.
\subsubsection{Comparison to pre-training models}
We compared BrainNPT with BrainNetTF~\cite{kan2022brain}, which achieved state-of-the-art performance in multiple brain functional network datasets. BrainNetTF is a Transformer based neural network, and it is different from our BrainNPT with different read-out layers. We implemented pre-training experiments with BrainNPT and BrainNetTF models using the same number of layers, 6 layers of Transformer. Moreover, BrainNPT and BrainNetTF were set with the same hyper-parameters except the read-out layers. The pre-training strategy RRP with 50\% probability to replace 50\% ROIs connectivity vectors on the data was used for two models.

Moreover, AdamW~\cite{loshchilov2017decoupled} was selected as the optimizer, and linear increasing warm-up and inverse square root decay strategy was selected as the scheduler for the learning rate~\cite{raffel2020exploring} during training. The learning rate was set as 1, with binary cross-entropy loss and 500 epochs for training. In experiments, we also randomly split 80\% of the dataset for training, 10\% for validation and 10\% for testing. Also, the performance are evaluated based on the average of 5 random runs on the test set with the standard deviation.

To evaluate the effectiveness of pre-training and fine-tuning, two pipelines were used: (i) pre-training on ABIDE I + REST-meta-MDD + HCP and fine-tuning on ABIDE II; (ii) pre-training on ABIDE I + HCP and fine-tuning on REST-meta-MDD. And we used accuracy and AUC as the metric for downstream classification experiments.

We compared the pre-trained BrainNPT model and the pre-trained BrainNetTF model on downstream classification tasks with the two pipelines respectively, and the results are shown in Table~\ref{tab:downstream}.
\begin{table}[h]
\centering
\caption{The performance of models with pre-training (mean±std).}
\label{tab:downstream}
\begin{tabular}{ccccc}
\hline
\multirow{2}{*}{Models}   & \multicolumn{2}{c}{ABIDE II}        & \multicolumn{2}{c}{REST-meta-MDD}               \\ \cline{2-5}   & Accuracy (\%)   & AUC (\%)      & Accuracy (\%)   & AUC (\%)     \\ \hline      
BrainNPT     & 62.50±6.24 & 64.35±5.05 & 65.42±3.21 & 68.46±3.17 \\
BrainNetTF   & 61.21±5.12          & 67.60±5.69   & 62.76±1.31          & 68.27±1.82  \\
BrainNetTF+pre-training   & 65.16±3.09 & 69.67±5.57  & 65.03±2.47 & 69.60±1.76\\ 
BrainNPT+pre-training & \textbf{71.25±3.42} & \textbf{71.54±11.94} & \textbf{66.67±3.58} & \textbf{70.55±3.48} \\ \hline
\end{tabular}
\end{table}
The results showed that the BrainNPT model with pre-training outperformed the BrainNetTF model with pre-training. The two models with pre-training achieved better performance than the models without pre-training. The BrainNPT model with pre-training outperformed the model without pre-training with the significant increase of more than 8\% in accuracy and more than 7\% in AUC.  

Compared with the performance of other twelve models in Table~\ref{tab:basic}, the BrainNPT model with pre-training achieved the best performance among all the metrics, which strongly indicated the effectiveness of our pre-training strategy RRP, with the capacity to enhance the comprehension of the structure of brain networks. Moreover, the improvement of the models on REST-meta-MDD was not better than the improvement on ABIDE II dataset, which might be caused by the smaller scale of pre-training datasets. That also indicated the importance of scale of pre-training dataset for Transformer-based models. Compared BrainNPT with BrianNetTF model, the specially designed <cls> embedding vector in BrainNPT might be sufficiently pre-trained by the RRP pre-training strategy, and that might have key impact on the performance of downstream classification tasks.

\subsubsection{Comparison to pre-training strategy}
\begin{figure*}[h]
    \centering    \includegraphics[width=0.3\linewidth]{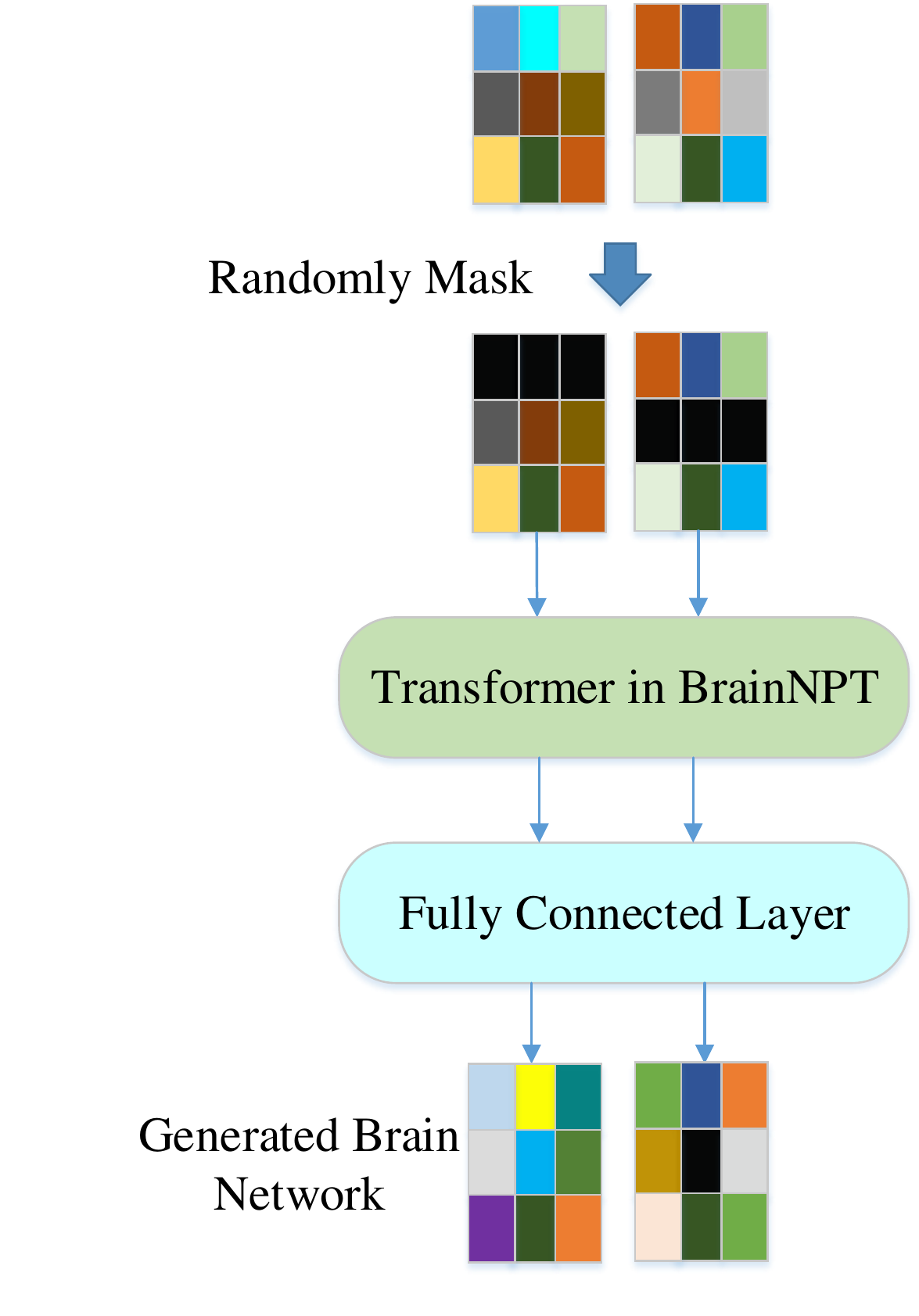}
    \caption{The framework of pre-training for MRM.}
    \label{fig:mrm}
\end{figure*}

To compare the pre-training strategies, we introduced a new pre-training strategy, namely Masked ROI Model (MRM), for brain funcitonal network learning. Inspired by the BERT pre-training strategy~\cite{devlin2018bert}, one of the most widely used pre-training strategies of Transformer, we assumed that randomly masking inputs and predicting the original masked values can help generate higher-level global representations and understand for the model. For brain networks, we randomly masked the functional connectivity vectors of some ROIs and utilized a pre-training model composed of Transformer and a fully connected layer to predict the original masked values. The framework of MRM is shown in Fig.~\ref{fig:mrm}.

We randomly masked functional connectivity vectors of 30\% ROIs of each input brain functional network for experiments and analysis. Referring to~\cite{clark2020electra}, the self-supervised task is to predict the values of entire original brain networks, which enables the model to learn more interaction information about brain networks. In order to keep the pre-training data and downstream fine-tuning data as consistent as possible, we did not set all of connectivity vectors to 0 while masking in experiments. Specifically, for the masked ROIs, we only randomly selected 80\% of the ROIs and set the entire connectivity vectors to 0, while randomly selecting and setting the remaining 10\% ROIs to random values from a normal distribution with the same range as the functional connectivity values, to further increase the difficulties of learning and similarity with the original data. Finally, the remaining 10\% of ROIs remained unchanged.

In the pre-training phase of the MRM, as the task is regression prediction, the <cls> classification embedding vector is not adopted. Instead, we forward the output of the Transformer part directly into a fully connected layer with an output shape of the same length as the functional connectivity vectors of ROIs. After employing a dropout layer to improve stability, the final output is compared with the original input brain network and performed supervised training using the mean-square-error (MSE) loss function. We pre-trained the models on ABIDE I, REST-meta-MDD, and HCP datasets, and fine-tuned the models for downstream classification on ABIDE II dataset. The results are shown in Table~\ref{tab:downstream}.

\begin{table}[h]
\centering
\caption{The performance of BrainNPT with MRM and RRP (mean±std).}
\label{tab:downstream}
\begin{tabular}{ccc}
\hline
Models       & Accuracy (\%)   & AUC (\%)         \\ \hline
BrainNPT     & 62.50±6.24 & 64.35±5.05  \\
BrainNPT+MRM & 66.67±4.29 & \textbf{71.73±6.26}  \\
BrainNPT+RRP & \textbf{71.25±3.42} & 71.54±11.94 \\ \hline
\end{tabular}
\end{table}
The results showed that the models with MRM and RRP pre-training both achieved better performance than the model without pre-training. The performance of MRM was not good as RRP strategy with lower accuracy in the experiments. The pre-training task of MRM was regression, while the pre-training task of RRP and downstream task were both classification. Accordingly, the optimization loss function of MRM used a mean square deviation, and the optimization loss function of RRP and downstream task used a cross entropy. Different optimization loss functions may affect the performance of the models. In addition, there was no <cls> classification embedding vector in the MRM model for pre-training, but the <cls> vector still needed to be trained in downstream fine-tuning classification tasks with small-scale data. 

\subsubsection{Analysis of training loss and testing accuracy in pre-training}
\begin{figure*}[h]
    \centering
    \includegraphics[width=0.9\linewidth]{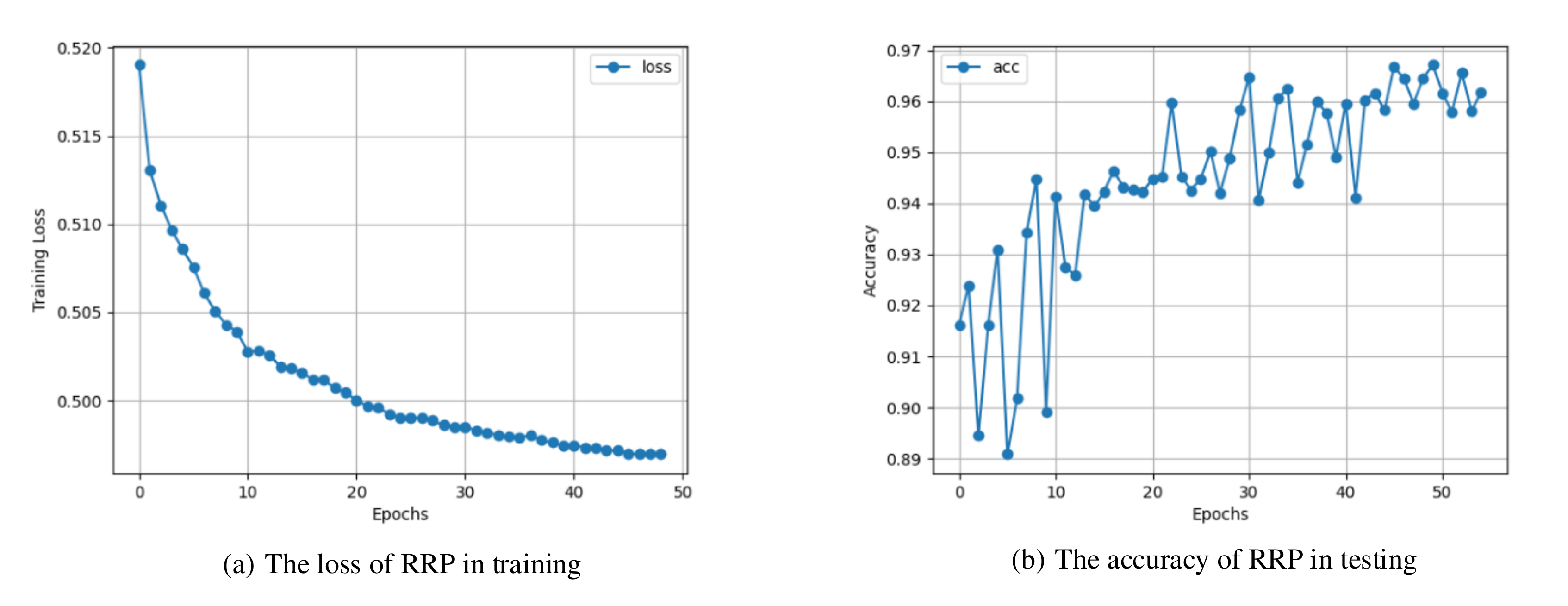}
    \caption{The training and testing evaluation of RRP model. The training losses tend to stabilize after about 45 rounds, and the accuracy of predicting replaced ROIs with RRP in testing is about 96\%.}
    \label{fig:pretrain_train}
\end{figure*}

In pre-training of BrainNPT model on ABIDE I, REST-meta-MDD, and HCP datasets, the loss of training and accuracy of testing are shown in Fig.~\ref{fig:pretrain_train}. The results showed that the training losses of RRP tended to stabilize after about 45 rounds. The classification accuracy of the RRP increased steadily with the increase of training rounds and finally oscillated up and down around 96\%. It indicated that RRP model converged with large-scale data for training, and effectively improved the performance of their pre-training tasks. 

\subsubsection{Influence of the replaced ratio of pre-training strategy}
To explore the influence of the ratio of replaced ROIs in RRP pre-training strategy, we conducted the replaced ratio comparative experiments, as shown in Table~\ref{tab:preratio}. We used ABIDE I as the pre-training dataset for pre-training. 95\% of the data were selected as pre-training sets, and 5\% of the data were used to evaluate the pre-training performance. Finally, the same settings of splitting training, validation and test sets were used on the ABIDE II dataset to evaluate the impact on downstream classification task. We compared the influence of replacing 30\%, 40\%, 50\%, and 60\% ROIs randomly.
\begin{table}[h]
\centering
\caption{Influence of replaced ratio in RRP (mean±std).}
\label{tab:preratio}
\begin{tabular}{ccc}
\hline
 Ratio & Accuracy (\%)       & AUC (\%)             \\ \hline

 60\%  & 63.19±5.09          & 64.46±7.69          \\
 50\%  & \textbf{64.58±4.73} & \textbf{66.84±6.28} \\
 40\%  & 63.89±8.71          & 63.77±7.96          \\
 30\%  & 63.19±7.67          & 61.08±7.65          \\ \hline
\end{tabular}
\end{table}
From Table~\ref{tab:preratio}, the results showed that the model with RRP achieved best performance when the replaced ratio reached 50\%. We could improve the performance of the model by increasing the replaced ratio when the replaced ratio is relatively low, but we could not improve the performance of the model when the replaced ratio reached over 50\%. That might be because continuously increasing the replaced ratio did not make a significant contribution to the learning of the model. 

\subsubsection{Influence of scale of pre-training data}
As the scale of public rs-fMRI datasets is generally limited with hundreds or thousands, and the scale of pre-training datasets for Transformer in NLP or CV can be more than millions, pre-training with unlabeled brain functional networks on Transformer remains challenging. 

\begin{figure*}[h]
    \centering
    \includegraphics[width=0.6\linewidth]{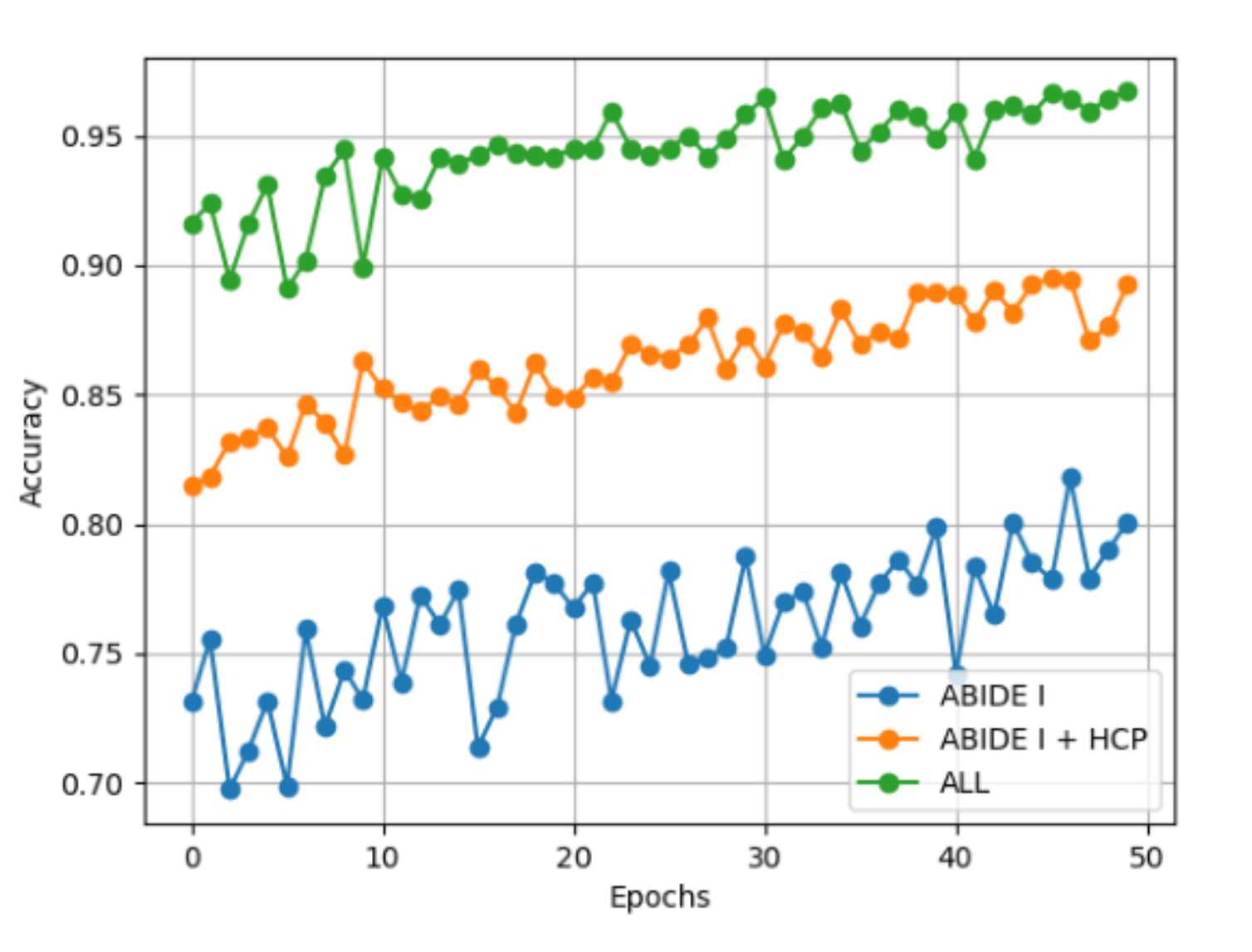}
    \caption{The accuracy of downstream classification with different scales of pre-training dataset.}
    \label{fig:all_train}
\end{figure*}

To evaluate the influence of scale of pre-training data, we compared the performance of downstream classification tasks on the different scale of pre-training data. We used three different pre-training dataset setting for RRP models: (i) ABIDE I dataset, (ii) ABIDE I and HCP datasets, (iii) All three datasets: ABIDE I, HCP, and REST-meta-MDD. The same ratio of training set, validation set and test set of the classification experiments was used to evaluate the performance of downstream classification on the ABIDE II dataset. Meanwhile, we conducted experiments with a unified replaced ratio of 50\% for RRP model. The performance of pre-training strategies under different scales of pre-training data are shown in Fig.~\ref{fig:all_train}.
\begin{table}[h]
\centering
\caption{The classification results on ABIDE II with different scale of pre-training datasets (mean±std).}
\label{tab:trainscale}
\begin{tabular}{ccc}
\hline
 Pre-training dataset & Accuracy (\%)       & AUC (\%)             \\ \hline
  ABIDE I              & 64.58±4.73          & 66.84±6.28           \\
                      ABIDE I + HCP        & 65.28±3.18          & 70.04±9.62           \\
                     ALL                  & \textbf{71.25±3.42} & \textbf{71.54±11.94} \\ \hline
\end{tabular}
\end{table}
The results from Fig.~\ref{fig:all_train} showed that under the same replaced ratio, the accuracy of replaced ROI prediction of RRP increased with the increasing scale of pre-training dataset, as about 80\% of the accuracy with only ABIDE I dataset, about 90\% of the accuracy with ABIDE I and HCP datasets, and almost 97\% of the accuracy with all three datasets. The results of downstream classification are shown in Table~\ref{tab:trainscale}, and that demonstrated the model achieved better performance with more pre-training data. 

\subsubsection{Influence of scale of fine-tuning dataset}
To evaluate the influence of scale of fine-tuning dataset, we used different scale of training data for fine-tuning BrainNPT on ABIDE II dataset. Selecting the models of best fine-tuning performance, pre-trained parameters of RRP, we conducted classification experiments with different scales of the training set, including 80\%, 70\%, 60\% and 50\% of the dataset for training, and 10\% of the dataset was set as the validation set. The results are shown in Table~\ref{tab:datascale}.
\begin{table}[h]
\centering
\caption{The performance of BrainNPT with different scale of training dataset (mean±std).}
\label{tab:datascale}
\begin{tabular}{cccc}
\hline
The ratio of training set & Pre-trained parameters loaded & Accuracy (\%)   & AUC (\%)        \\ \hline
\multirow{2}{*}{80\%}     & No                            & 62.50±6.24 & 64.35±5.05 \\
                          & Yes                           & 70.83±3.10 & 71.68±4.34 \\ \hline
\multirow{2}{*}{70\%}     & No                            & 60.42±4.15 & 64.84±6.39 \\
                          & Yes                           & 63.19±7.54 & 63.79±5.95 \\ \hline
\multirow{2}{*}{60\%}     & No                            & 58.85±4.24 & 62.86±5.78 \\
                          & Yes                           & 60.42±6.32 & 62.82±5.93 \\ \hline
\multirow{2}{*}{50\%}     & No                            & 56.94±5.53 & 58.92±7.53 \\
                          & Yes                           & 60.94±5.82 & 63.51±6.79 \\ \hline
\end{tabular}
\end{table}
The results showed the performance of the models sharply dropped under the same settings when the scale of training set was decreased. The accuracy and AUC were finally stabilized around 57\% and 60\%, respectively. It indicated that the reduction of training data had a great impact on the performance of the model. The BrainNPT model with pre-training still maintained a high level of performance and was significantly higher than the models without pre-training, especially with only 50\% of the training data, the classification accuracy still remained at 60\%, compared with only 56.94\% for the BrainNPT model without pre-training. These demonstrated that the pre-training would significantly enhance the performance of downstream classification tasks even with limited labeled data.

\subsection{Model interpretation}
\label{sec:interpret}
\subsubsection{Key ROIs in classification model}
Based on the best performance model of BrainNPT with RRP pre-training, the interpretation experiments were conducted to explore the most relevant ROIs for the classification. With the trained BrainNPT model, for any given brain network, we can obtain the relevance scores of corresponding ROI with LRP methods. Specifically, with all relevance scores obtained, Min-Max standardization was performed to scale the scores in the range of (0,1). The operation of Min-Max standardization is shown as the follows.
\begin{equation}
\bar C{\rm{ = }}\frac{{C{\rm{ - }}{{\rm{C}}_{\min }}}}{{{{\rm{C}}_{\max }}{\rm{ - }}{{\rm{C}}_{\min }}}}
\label{eq:standard}
\end{equation}
Where ${{\rm{C}}_{\max }}$ and ${{\rm{C}}_{\min }}$ represents the highest and the lowest relevance scores. 

\begin{figure*}[h]
    \centering
    \includegraphics[width=0.8\linewidth]{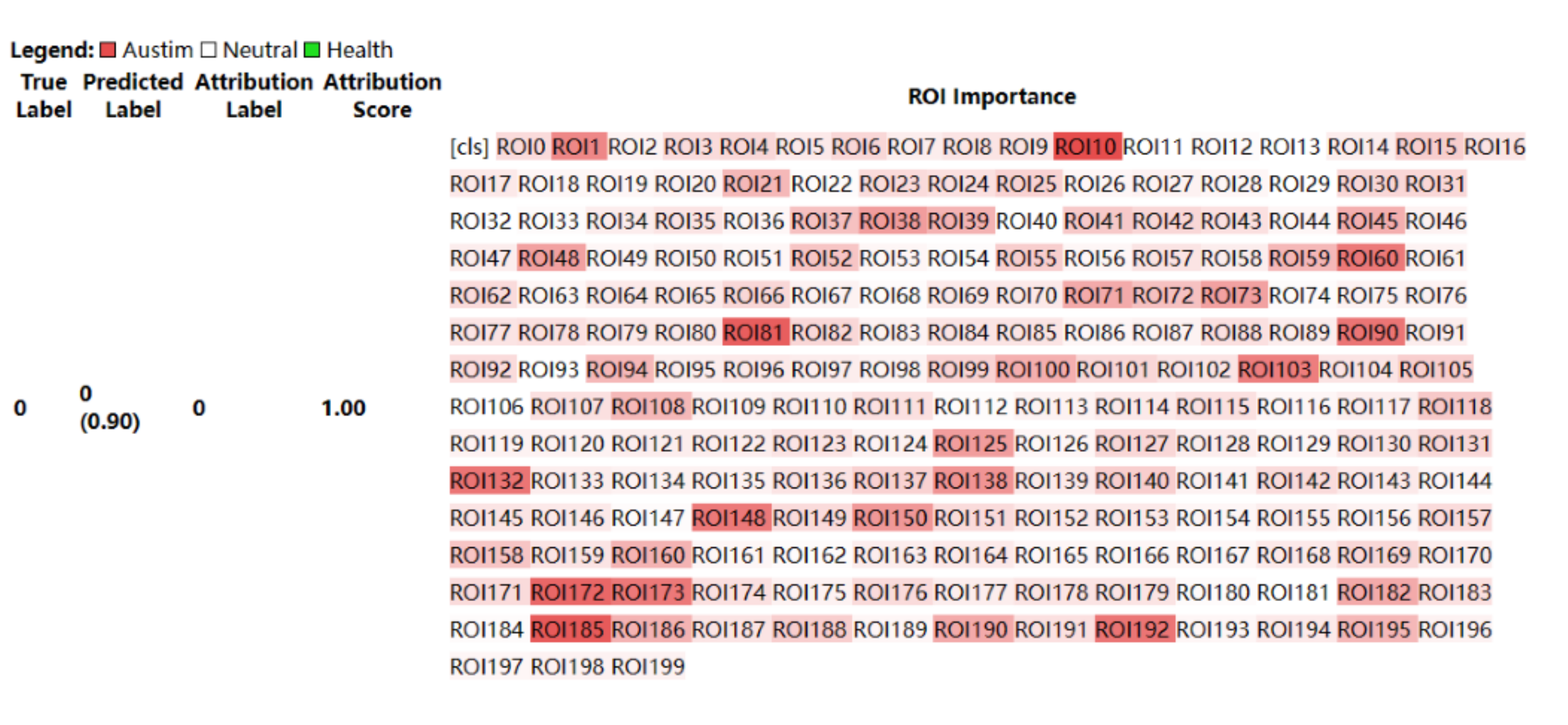}
    \caption{Visualization result for a subject predicted to be Autism. The deeper color of ROI ID represents the higher relevance of the ROI with Autism.}
    \label{fig:asd}
\end{figure*}

\begin{figure*}[h]
    \centering
    \includegraphics[width=0.8\linewidth]{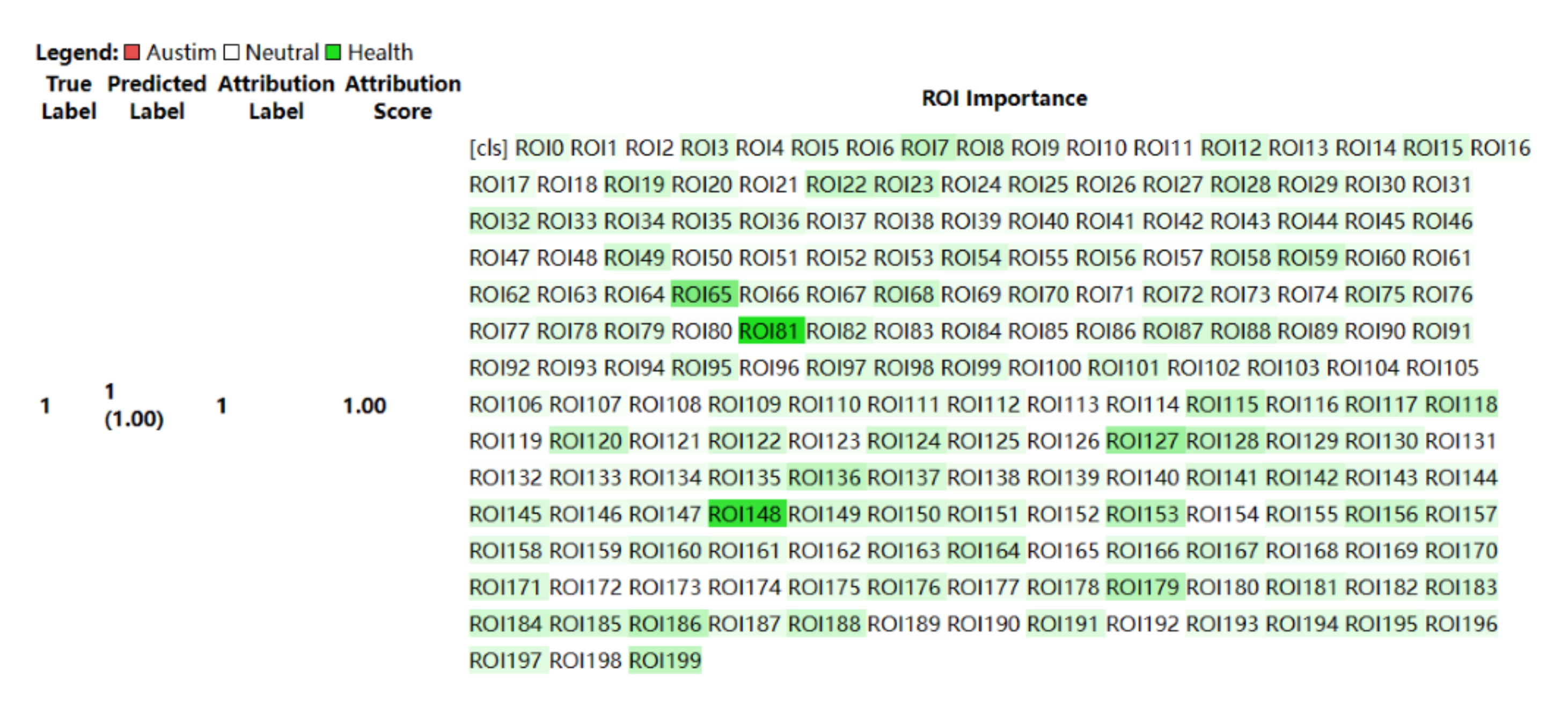}
    \caption{Visualization result for a subject predicted to be healthy control. The deeper color of ROI ID represents the higher relevance of the ROI with health.}
    \label{fig:hc}
\end{figure*}

Scaled relevance scores can be visualized to observe the most related ROIs for a predicted class of each subject, as shown in Fig.~\ref{fig:asd} and Fig.~\ref{fig:hc}, where the deeper color representing the higher relevance.

To obtain the most important ROIs of predicting ASD by BrainNPT model on ABIDE II, a frequency based methods~\cite{hu2020interpretable} was implemented for evaluating the importance of ROIs in the model. For each subject, the $K$ ROIs with the highest relevance scores were added into the ROI subset ${F_K}$. After counting all subjects, the ROIs in the subset ${F_K}$ with the frequencies more than 60\% of the number of the subjects were considered as the most important ROIs for predicting autism subjects. To obtain the top 3 key ROIs for predicting autism subjects, we gradually raised the number of $K$ to 84 until 3 ROIs satisfied the condition of the most important ROIs described above. The top 3 key ROIs were the 191st ROI, the 82nd ROI and the 2nd ROI from Craddock200 atlas. 

\subsubsection{Analysis of the top key ROIs}
Since the ROIs from Craddock200 atlas are clustering ROIs, we analyzed these ROIs according to their corresponding ROIs from AAL brain atlas~\cite{tzourio2002automated}, which was most commonly used in brain imaging analysis. The top 3 ROIs involved five ROIs in AAL brain atlas, namely left middle temporal gyrus (Temporal\_Mid\_L), left angular gyrus (Angular\_L), left middle occipital gyrus (Occipital\_Mid\_L), left superior frontal gyrus (Frontal\_Sup\_L) and left middle frontal gyrus (Frontal\_Mid\_L). 

Our interpretation results are consistent with previous studies about ASD. For example, the angular gyrus has been shown to be highly correlated with reading, text comprehension, and attention~\cite{seghier2013angular}, and the left angular gyrus has also been shown to be associated with social interaction defects~\cite{lai2014autism}. Moreover, as a key area of the brain for processing social networks, the middle temporal gyrus has been shown to be highly correlated with autism~\cite{bachevalier1994medial}, and recent research has further verified that there was significant functional connectivity variation in the left middle temporal gyrus in autistic patients~\cite{xu2020specific}. In addition, the middle occipital gyrus is a significant brain region responsible for object recognition and often participates in facial recognition and emotion recognition tasks~\cite{banich2018cognitive}. It has also been shown that there were differences in functional connectivity between patients with autism and healthy controls with performing facial expression recognition tasks~\cite{pelphrey2007perception}. As the superior frontal gyrus is an important part of the default mode network~\cite{raichle2015brain} and plays a key role in social behavior, it has also been found that there were functional connectivity damages in autistic patients~\cite{monk2009abnormalities}. Additionally, the middle frontal gyrus has been discovered to be related to the pathophysiology of autism in neurocognitive studies~\cite{zikopoulos2013altered,barendse2013working,carper2005localized}, with abnormal cortical volume and connectivity, resulting in working memory processing deficiency and other abnormalities.


\section{Conclusions}
\label{sec:conclusion}
This paper proposed a Transformer-based neural network and a pre-training framework for brain functional network classification, which leveraged pre-training to capture the intrinsic structure information from unlabeled brain network data. A learnable <cls> token as a classification embedding vector was proposed for the Transformer to capture the representation of brain network, and replaced ROI prediction strategy was proposed for pre-training. The results of experiments demonstrated the BrainNPT model with pre-training outperformed the state-of-the-art models. Moreover, the pre-training strategies and the influence of the parameters of the model were further analyzed, and the trained BrainNPT model was interpreted by LRP.

In the future, BrainNPT could be improved with deeper layers of Transformer and more unlabeled brain functional network data, and it could be used as the pre-training model for further brain network analysis, such as identifying the subtypes of mental disorders, and characterizing brain functional connectivity and their variability in healthy adults.
~\\\\[3pt]
\textbf{CRediT authorship contribution statement} 
\newline \textbf{Jinlong Hu:} Conceptualization, Resources, Methodology, Investigation, Writing – original draft, Writing – review \& editing, Supervision, Project administration, Funding acquisition. \textbf{Yangmin Huang:} Conceptualization, Methodology, Investigation, Writing – original draft, Writing – review \& editing. \textbf{Nan Wang:} Writing – review \& editing. \textbf{Shoubin Dong:} Conceptualization, Methodology, Resources, Funding acquisition.
~\\\\[3pt]
\textbf{Declaration of Competing Interest}
\newline The authors declare no competing interests.
~\\\\[3pt]
\textbf{Acknowledgements and Funding}
\newline This research was supported by the Natural Science Foundation of Guangdong Province [2021A1515011942], and the Innovation Fund of Introduced High-end Scientific Research Institutions of Zhongshan [2019AG031]. We would like to thank Dr. Xiaoxiao Li of the University of British Columbia for sharing her comments to improve this work.
~\\\\[3pt]
\newline 

\bibliographystyle{unsrt}  
\bibliography{references}  


\end{CJK}
\end{document}